\documentclass[a4paper,conference]{IEEEtran}

\usepackage{cite}      
\usepackage{graphicx}  
\usepackage{subfigure} 
\usepackage{url}
\usepackage{amsmath}   
\usepackage{array}
\usepackage[makeroom]{cancel}

\usepackage{algpseudocode}
\usepackage{algorithm}
\usepackage{algorithmicx}
\usepackage[dvipsnames]{xcolor}
\usepackage{mathrsfs}
\usepackage{xspace}
\usepackage{doi}
\usepackage{soul}

\usepackage{hyperref}
\hypersetup{
    colorlinks=true,
    linkcolor=blue,
    filecolor=magenta,      
    urlcolor=cyan,
}

\graphicspath{{plots/}}

\newcommand{\edit}[2]{#2}  

\usepackage{fancyheadings}
\newcommand{\flecsph}[1]{{\textsc{FleCSPH}\xspace#1}}

\begin{document}

\title{Modeling Neutron Star Oscillations in a Fixed General Relativistic
Background Including Solid Crust Dynamics}

\author{
\IEEEauthorblockN{Bing-Jyun Tsao, Irina Sagert, Oleg Korobkin, Ingo Tews, Hyun Lim, Gary Dilts, Julien Loiseau}
\IEEEauthorblockA{Los Alamos National Laboratory\\
Los Alamos, NM 87544\\
btsao@lanl.gov}}

\maketitle

\begin{abstract}
Measurements of the gravitational-wave signals from neutron star mergers allow scientists to learn about the interior of neutron stars and the properties of dense nuclear matter. The study of neutron star mergers is usually performed with computational fluid dynamics codes, mostly in Eulerian but also in Lagrangian formulation such as smoothed particle hydrodynamics (SPH). Codes include our best knowledge of nuclear matter in the form of an equation of state as well as effects of general relativity (GR). However, one important aspect of neutron stars is usually ignored: the solid nature of their crust. The solid matter in the crust is the strongest material known in nature which could lead to a multitude of possible observational effects that have not been studied in dynamical simulations yet. The crust could change the way a neutron star deforms during a merger, leaving an imprint in the gravitational wave signal. It could even shatter during the inspiral, producing a potentially observable electromagnetic signal. Here, we present a first study of the dynamical behavior of neutron stars with a solid crust and fixed GR background with \flecsph. \flecsph is a general-purpose SPH code, developed at Los Alamos National Laboratory. It features an efficient algorithm for gravitational interactions via the Fast Multipole Method, which, together with the implemented nuclear equation of state, makes it appropriate for astrophysical applications. The solid material dynamics is described via the elastic-perfectly plastic model with maximum-strain breaking. Despite its simplicity, the model reproduces the stress-strain curve of crustal material as extracted from microphysical simulations very well.
We present first tests of our implementation via simulations of neutron star oscillations and give an outlook on our study of the dynamical behavior of the solid crust in neutron star merger events.

\end{abstract}

\section{Introduction}
Neutron stars are astrophysical laboratories for dense nuclear matter~\cite{Lattimer:2006xb, Gandolfi2015}.
Born from the explosion of massive stars in  core-collapse supernovae, neutron stars can have masses of up to about twice the mass of the sun but a radius of only ca. 12km~\cite{Dietrich:2020efo}. 
Matter in the center of neutron stars is so dense that it exceeds the density in atomic nuclei up to several times. 
Information on the equation of state (EoS) of dense nuclear matter can be extracted from  measurements of the stars' masses and radii. 

Mass measurements are possible in binary systems of two compact stars where at least one is a pulsar, i.e. a neutron star that emits synchrotron radiation along its magnetic poles.
Such a star appears as a very regularly pulsating electromagnetic source if Earth lies in the path of the radiation beam. 
By timing the pulsar signal, mass determination is achieved by measuring the so-called post-Keplerian parameters. 
If the companion is a white dwarf, spectroscopic measurements of the white dwarf atmosphere can also be used~\cite{Ozel2016}. 
Combined radius and mass measurements can be achieved, for example, for neutron stars in low-mass X-ray binaries where the star accretes hydrogen or helium from a companion, e.g. a white dwarf or a main-sequence star~\cite{naettilae2017}.
When accreted matter on the neutron star surface ignites, it emits an X-ray burst and temporarily levitates the photosphere.
The photosphere eventually touches down on the neutron star surface. 
Measuring the touchdown flux and assuming that it equals the Eddington flux allows the extraction of the neutron star mass and radius.  
In 2017, the NICER observatory started to measure the compactness, i.e. ratio of mass and radius, of non-accreting millisecond pulsars from rotating hot-spots~\cite{Bogdanov2019} using X-ray pulse-profile modeling and provided first data in late 2019~\cite{Riley:2019yda, MillerLamb2019}.
Since then, the NICER collaboration also measured the radius of the heaviest neutron star known to date~\cite{Miller:2021qha, Riley:2021pdl}.

With the first observation of gravitational waves (GW) from a neutron star merger in 2017, GW170817, by the Advanced LIGO/VIRGO collaboration, a new method became available to gather information on dense nuclear matter~\cite{Abbott2017g,ligo16}. 
The so-called tidal deformability, a quantification of a star's quadrupolar deformation under the influence of the gravitational field from the companion, is dependent on the nuclear matter EoS and can be extracted from the measured GW signal~\cite{ligo17a, De2018}.
In addition to the GW signal, GW170817 was also accompanied by electromagnetic counterparts: a
gamma-ray burst, an afterglow, and a so-called kilonova~\cite{Abbott17m}.
The first GW detection of a binary neutron star (BNS) merger has been followed by one more confirmed event in LIGO's 3rd observational run, GW190425~\cite{ligo20}, and several potential candidates involving at least one neutron star.
Unlike GW170817, none of these more recent events were accompanied by an electromagnetic counterpart.
Nevertheless, it is expected that many future merger events will be multimessenger, i.e. have an electromagnetic counterpart: a gamma-ray burst, an afterglow, or a so-called kilonova, similar to GW170817.

With more detections in the future and more precise measurements of GW signals we will be able to extract information not only on the bulk properties of neutron stars but also possibly on the solid neutron star crust. 
The latter covers the outermost layer of the neutron star and is formed by a lattice of increasingly neutron-rich nuclei, spanning around (1-2)km in depth~\cite{Chamel2008}. 
The so-called outer crust is a Coulomb crystal of neutron-rich nuclei surrounded by a degenerate electron gas. 
The inner crust is formed by a lattice of progressively heavier and more neutron-rich nuclei, surrounded additionally by a gas of free neutrons. 
Neutron star crust material is the strongest material in Nature, about 10$^{10}$ times stronger than terrestrial engineering materials with a breaking strain of about 0.1~\cite{Horowitz2009}. 
Molecular dynamics simulations of material in the neutron star outer crust have shown that, under shear deformation, it behaves like a brittle solid, namely, deforming elastically before undergoing abrupt failure. 
Due to the large pressure in the crust, the material fails collectively at large strains rather than yielding or fracturing at low strains. 
For the inner crust, the solid behavior might be more complex, especially in the presence of nuclear pasta, including plasticity~\cite{Caplan2018}.

The large strength of the neutron star crust may lead to several observable effects during neutron star mergers: 1) elastic deformation, elastic-to-plastic transitions but also continuous breaking during the merger due to tidal deformation can lead to a phase shift in the GW waveform which could be detectable in current and future GW observatories~\cite{Abbott2017g, xu17, gittins20, pan20}; or 2) excitation of interface mode (or i-mode) oscillations between the neutron star crust and core during the inspiral could result in crust shattering which may produce observable electromagnetic signals before mergers, potentially the only detectable counterpart to the GWs from BH-NS mergers~\cite{tsang12, Neill2021}. 
This phenomenon could also explain so-called Gamma-Ray-Burst (GRB) precursors that might have been observed for some GRBs~\cite{troja10} \edit{}{(but see also~\cite{minaev18}).}

\edit{To understand the dynamics of BNS, various methods in numerical relativity are aimed to solve the Euler and Einstein equations for a BNS system. Baumgarte et al. presented a formulation to solve Einstein's equations for binary neutron stars in quasi-equilibrium.
Additionally, grid-based codes such as the Spectral Einstein Code (SpEC) and the Einstein Toolkit  have been developed for general relativistic hydrodynamics that have been used for compact objects simulations.
To study the system of binary neutron stars including the deformation of the stars and the fracture of the solid crust, here we use smoothed particle hydrodynamics (SPH) to solve the general relativistic fluid equations.}{} \edit{}{Simulating the dynamics of BNS mergers will require numerical relativity with fully coupled spacetime metric and fluid equations.~\cite{rosswog15}
As of now, several high-fidelity grid-based relativistic codes have been developed to simulate neutron star binaries, including the Spectral Einstein Code (SpEC)\cite{spec}, the Einstein Toolkit~\cite{einsteintoolkit}, the wavelet-adaptive mesh refinement code DendroGR~\cite{Neilsen19}, etc.
In the past, the method of smoothed-particle hydrodynamics (SPH) has been successfully used to model astrophysical BNS mergers only in Newtonian or conformally-flat approximations (e.g.~\cite{rosswog13, bauswein13}).
However, recently SPH has also made an entry into the field of numerical relativity, with debut work  of
Rosswog \& Diener~\cite{rosswog20}.}

The use of SPH has the following advantages: 1) the Lagrangian fluid equations have a simpler form and there is no need to transform from conservative quantities to primitive quantities such as density or velocity (which is often used in grid-based codes where non-linear equations with primitive variables as unknowns have to be solved), 2) \edit{properties like mass, momentum, and energy are conserved}{properties of the particles such as their mass or composition are trivially (and exactly) advected with the flow},
and 3) the use of particles as interpolation points automatically sets up adaptive refinement for numerical accuracy, and it is easier to handle vacuum on the surface of the star (which often poses instability in grid-based codes that requires an artificial atmosphere to resolve the issue). 
Finally, while somewhat novel for astrophysical applications, fluid-structure interactions and interactions with static boundaries have been widely explored with SPH for engineering and animation studies~\cite{Sun2021, Liu2019, Hu2017, Akinci2013, Adami2012, Keiser2005, Mueller2003}.

In the following sections, we will discuss the formalism and implementation of FleCSPH, a multiphysics FleCsible parallel computational infrastructure for SPH, to simulate relativistic fluids and the solid crust with an end goal to produce binary neutron star simulations with crust dynamics and, potentially, breaking.

\subsection{FleCSPH}
\textsc{FleCSPH}\footnote{https://github.com/laristra/flecsph} is an open-sourced SPH code developed at Los Alamos National Laboratory (LANL) that supports multi-physics simulations~\cite{loiseau20}.
\flecsph is based on the framework provided by the Flexible Computational Science Infrastructure (\textsc{FleCSI}~\cite{charest17}). 
\textsc{FleCSI}\footnote{https://github.com/laristra/flecsi} is a task-based runtime abstraction layer that provides a seamless programming model for distributed-memory tasks (Legion), and fine-grained data-parallel kernels (Kokkos), with several core topology types that can be statically specialized to support a variety of applied methods. 
When using \textsc{FleCSI}, \flecsph has the potential to separate the application implementation from the details of machine architecture. 
\flecsph is designed as a general-purpose code. 
Its current main applications lie in astrophysical problems, with a focus on compact stars (white dwarfs and neutron stars) and binary mergers. 
The code contains different SPH kernels, astrophysical and material equations of state in tabulated and analytic form, Newtonian gravity via N-body calculations and the Fast-Multipole Method (FMM), as well as external potentials for the implementation of boundary conditions. 
The recent additions to \flecsph's features are a fixed general-relativistic background for static and rotating neutron stars and an elastic-perfectly plastic strength model which we apply to simulate neutron stars with a solid crust. 

\subsection{Solid SPH Equations}
\edit{For material strength, FleCSPH currently includes}{Our material strength model includes} equations for elastic solids, von-Mises plasticity and a maximum-strain damage model. 
The corresponding SPH equations for the conservation of mass, momentum, and energy, as well as the time evolution of the deviatoric stress tensor $S^{ij}$ at the position of particle $a$ are: 
\begin{equation}
    \frac{d \rho_a}{dt} = \rho_a \sum_b \frac{m_b}{\rho_b} v_{ab}^i \frac{\partial W_{ab}}{\partial x_a^i}\,,
    \label{solid1}
\end{equation}
\begin{align}
    \frac{dv_a^i}{dt} &= - \sum_b m_b \left[\left(\frac{P_a}{\rho_{a}^2} + \frac{P_b}{\rho_{b}^2} + \Pi_{ab} \right) \frac{\partial W_{ab}}{\partial x_a^i} \right. \nonumber \\
    & \left. - \left(\frac{S_a^{ij}}{\rho_{a}^2} + \frac{S_b^{ij}}{\rho_{b}^2} \right)  \frac{\partial W_{ab}}{\partial x_a^j} \right]\,,
    \label{solid2}
\end{align}
\begin{equation}
    \frac{du_a}{dt} = \sum_b m_b \left( \frac{P_a}{\rho_a^2} + \frac{\Pi_{ab}}{2} \right) v_{ab}^i \frac{\partial W_{ab}}{\partial x_a^i} + \frac{S_a^{ij}}{\rho_a} \dot{\epsilon}^{ij} \,,
    \label{solid3}
\end{equation}
\begin{equation}
  \frac{d S^{ij}}{dt} = 2\mu (\dot{\epsilon}^{ij} - \frac{1}{3} \delta^{ij} \dot{\epsilon}^{kk}) + S^{il} \dot{R}^{lj} - \dot{R}^{il}S^{lj}\,,
  \label{solid4}
\end{equation}
where the summation is over particles $b$ within the support radius of particle $a$. 
This implementation of SPH particles uses the shear modulus $\mu$, the viscosity $\Pi_{ab}$, and the smoothing kernel $W_{ab}$. Here, $W_{ab} = W(|\vec{r}_a - \vec{r}_b|, h_{ab})$ is characterized by the smoothing length $h_{ab} = (h_a + h_b)/2$, and $v_{ab} = v_a - v_b$.
Furthermore, in the above equations, while Latin letters from the beginning of the alphabet ($a$, $b$, ...) are particle labels, letters from the middle of the alphabet ($i$, $j$, ...) are used as spatial indices.
Also, repeated indices are implicitly summed over (Einstein's rule).
The strain rate and rotation rate tensors in Eqs.~(\ref{solid3}) and~(\ref{solid4}) are given by:
\begin{align}
    \dot{\epsilon}^{ij} &= \frac{1}{2\rho_a} \sum_b m_b \left[ v_{ba}^i \frac{\partial W_{ab}}{\partial x_a^j} + v_{ba}^j \frac{\partial W_{ab}}{\partial x_a^i}  \right]\,,
    \label{solid5}\\
    \dot{R}^{ij} &= \frac{1}{2\rho_a} \sum_b m_b \left[ v_{ba}^i \frac{\partial W_{ab}}{\partial x_a^j} - v_{ba}^j \frac{\partial W_{ab}}{\partial x_a^i}  \right]\,,
    \label{solid6}
\end{align}
respectively. 
The von-Mises yield criterion is implemented by scaling the deviatoric stress tensor according to:
\begin{align}
    S^{ij} &\rightarrow f S^{ij}, \:\: f = \mathrm{min} \left[\frac{Y_0^2}{\sigma_v^2}, 1 \right], \:\: 
    \sigma_v = \sqrt{\frac{3}{2} S^{ij} S^{ij}},
\end{align}
i.e. when the von-Mises stress $\sigma_v$ reaches or exceeds the yield stress $Y_0$.
In the maximum-strain damage model, failure occurs once the breaking strain is reached. 
For this, we compute the local scalar strain from the maximum tensile stress $\sigma_\mathrm{max} = \mathrm{max} \left[\sigma_1, \sigma_2, \sigma_3 \right]$ where $\sigma_\gamma$ are the principal stresses. 
The local strain is then given by 
\begin{align}
    \epsilon = \frac{\sigma_\mathrm{max}}{E}\,, \:\:\: E = \frac{9B\mu}{3B+\mu}\,,
\end{align}
with the shear modulus $\mu$ and bulk modulus $B$. 
For our planned work, the maximum-strain damage model will not be used for post-damage flow where it might not be sufficiently accurate.
Instead, we will only use it to determine onset of crust-breaking during a neutron-star merger event. \\
The solid material modeling capabilities of \flecsph have been verified using various standard tests including the colliding rubber rings, the Taylor anvil, and the Verney implosion with good agreement to published work~\cite{sagert21}.

\subsection{Fixed metric background for general relativistic fluids}
\edit{To implement a fixed GR background into FleCSPH, we apply the Cowling approximation, where}{In a fixed GR background case (the so-called Cowling approximation),} the dynamics of spacetime is ignored such that particles are moving in a time-independent gravitational \edit{potential}{field}.
This approach is a good approximation to simulate relativistic objects near equilibrium, \edit{which includes}{such as} isolated static or rotating neutron stars\edit{ as well as}{, or} neutron stars in an external quadrupolar field that is e.g. sourced by a companion in a binary system. 
However, for highly dynamical simulations where two or more relativistic bodies are involved, such as BNS shortly before the merger, fully dynamical GR implementations are needed.

For the rest of the paper, in addition to the previously mentioned \edit{particle labels}{notation}, we will apply the following \edit{notation}{conventions}: (a) geometric units ($G=c=M_{\odot}=1$); (b) the ``East Coast'' $(-+++)$ metric signature; (c) Greek letters \edit{will be assigned to}{for} spacetime indices.
For the relativistic implementation of SPH, we apply the 3+1 decomposition where the metric is written as
\begin{eqnarray}
ds^2 = -\alpha^2 dt^2 + \gamma_{ij}(dx^i+\beta^i dt)(dx^j +\beta^j dt)\,,
\end{eqnarray}
with $\alpha, \beta^i, \gamma_{ij}$ being the lapse, shift vector, and the spatial metric, respectively.
\edit{This formulation assumes}{We simulate} a perfect fluid with the stress-energy tensor
\begin{eqnarray}
T^{\mu\nu} = \rho h u^\mu u^\nu + P g^{\mu\nu}\,,
\end{eqnarray}
where $h$ is the specific enthalpy, $u^\alpha$ is the 4-velocity, and $g_{\mu\nu}$ is the metric.
\edit{The equation of motion for a relativistic fluid in Lagrangian form is given by}
{Conservation of $T^{\mu\nu}$ results in the following expression for fluid trajectories (as derived in Tejeda et al.~\cite{Tejeda17})}:
\begin{align}
\frac{d^2x^i}{dt^2} &= 
\frac{d\dot{x}^i}{dt} =
- (g^{i \lambda} - \dot{x}^i g^{0\lambda}) \nonumber\\
&\times \left[ \frac{1}{\tilde W \rho h}\frac{\partial P}{\partial x^\lambda} +  \left(\frac{\partial g_{\mu\lambda}}{\partial x^\nu} - \frac{1}{2} \frac{\partial g_{\mu\nu}}{\partial x^\lambda}\right) \dot{x}^\mu \dot{x}^\nu 
\right]\,,
\end{align}
where the generalized Lorentz factor $\tilde W$ is:
\begin{equation}
    \tilde W = (-g_{\mu\nu} \dot{x}^\mu \dot{x}^\nu)^{-1/2}\,.
\end{equation}
We ensure conservation of baryon number by assigning each particle an equal baryonic mass $m_b$ instead of assigning their rest mass.
The rest mass density in the frame of the fluid is recovered from the so-called ``computational density'',
\begin{eqnarray}
  D = \rho \sqrt{\gamma}\;W\,,
\end{eqnarray}
where $\gamma$ is the determinant of $\gamma_{ij}$ and $W$ is the usual Lorentz factor $W \equiv (1-v^i v_i)^{-1/2}$ that involves the fluid 3-velocity,
\begin{eqnarray}
v^i = \frac{1}{\alpha}\left(\frac{dx^i}{dt} + \beta^i\right)\,,
\end{eqnarray}
in the frame of an observer with a worldline normal to $t=\text{const}$.
The computational density $D$ is recovered from the SPH particle distribution using the standard formula.

\section{Numerical implementation}
\label{sec:num}
\subsection{Setting up Neutron Stars in FleCSPH}
To simulate a neutron star, we first require a nuclear high-density EoS to describe matter in its interior. 
\flecsph has several analytic and tabulated EoS options for astrophysics, including \edit{the}{an} ideal gas and cold white dwarf models as well as tabulated finite-temperature nuclear astrophysics EoSs.
For nuclear matter in isolated and binary neutron stars, polytropic EoSs of the form
\begin{equation}
    P = K \rho^\Gamma\,,
    \label{poly_eos}
\end{equation}
are a good approximation.
Here, $\rho$ is the baryon mass density, and $K$ and $\Gamma$ are parameters.
To refine the modeling of the crust and the core several polytropic models with different values for $K$ and $\Gamma$ can be connected to a so-called piecewise polytropic EoS~\cite{Read2009}.

In \flecsph, we can initialize different geometric objects like cuboids, rings, cylinders, and spheres using rectangle, hexagonal closest packed (hcp), face-centered cubic (fcc), and random particle placing. 
For neutron stars, we use an icosahedral grid where particles are distributed in concentric spherical shells. 
The radial distribution of particles depends on the neutron star density profile, which can be determined from a given central density $\rho_c$ and EoS by solving the Tolman–Oppenheimer–Volkoff (TOV) stellar structure equations: 
\begin{align}
    \frac{ds}{d\theta} =& -2 n \frac{s^{3/2}}{m\theta}\left(\frac{dP}{d\rho}\right)\left(1 + u + \frac{P}{\rho}\right)^{-1} \nonumber\\
    &\times \left(1+\frac{4\pi r^3 P}{m}\right)^{-1} \left(1 - \frac{2m}{r} \right)\,, \\
    \frac{dm}{d\theta} =& 2\pi s^{1/2}\rho\frac{ds}{d\theta}\left(1+ u\right)\,,
\end{align}
where $s = r^2$, the dimensionless density $\theta \in [0,1]$ satisfies
\begin{align}
    \rho = \rho_c \theta^n\,, \:\:\: n =\frac{1}{\Gamma -1}\,,
\end{align}
and the specific internal energy density $u$ is
\begin{align}
    u = \int \frac{P}{\rho^2} d\rho\,.
\end{align}

To avoid numerical artifacts which can occur when particles are placed on a regular lattice,
we perturb the particle positions from the initial icosahedral placing. 
This random perturbation is followed by a relaxation step where particles can find the energetically most favorable configuration. 
Particle relaxation can be done in an external potential that uses the desired density profile $\rho(r)$ as \edit{}{an} input.
In equilibrium, the momentum equation 
\begin{equation}
    \frac{d\vec{v}}{dt} = - \frac{1}{\rho} \nabla P + \vec{g}_\mathrm{eff}
\end{equation}
reduces to $\frac{1}{\rho} \nabla P  = \vec{g}_\mathrm{eff}$,
where $g_\mathrm{eff}$ is the effective external acceleration. 
The latter can be represented as a gradient of the external potential. Using e.g. a polytropic EoS,
\begin{align}
    - \nabla \phi_\mathrm{eff} = \vec{g}_\mathrm{eff} 
    = K \: \Gamma \rho^{\Gamma -2} \nabla \rho\,.
\end{align}
With that, the required external potential for a given \edit{radial}{density} profile is given by:
\begin{equation}
    \phi_\mathrm{eff} = - \frac{K \Gamma \rho^{\Gamma -1}}{\Gamma -1}\,.
\end{equation}
In addition, by adding potential walls, we can confine the particles in the effective potential.

Time steps during the relaxation process are largely dependent on the speed of sound $c_s$ in the star. 
For large density variations, and therefore differences in $c_s$, between e.g. the star's center and surface this can result in long relaxation times.  
Here, we can apply a trick which results in a mostly uniform sound speed throughout the entire star. 
For a polytropic EoS, the sound speed in the Newtonian approximation is: 
\begin{align}
    c_s = \sqrt{\frac{dP}{d\rho}} = \sqrt{\Gamma K \rho^{\Gamma - 1}}\,.
\end{align}
\edit{We can see that by}{By} using a value for $\Gamma$ which is close to unity, we can \edit{}{largely} remove the dependence of $c_s$ on the density. 
For relativistic stars, the sound speed is 
\begin{align}
    c_s = \sqrt{\frac{dP}{d\epsilon}}\,, \:\:\: d\epsilon = \frac{\epsilon + P}{\rho} \: d\rho\,,
\end{align}
with the energy density $\epsilon = (1 + u)\rho$.
This results in 
\begin{align}
    c_s = \sqrt{\frac{\Gamma K \rho^{\Gamma -1}}{1 + u + P/\rho}}\,, \:\:\: u = \frac{K \rho^{\Gamma - 1}}{\Gamma - 1}\,.
\end{align}
\edit{We can see that it is usually a good choice to}{During the relaxation process, we} set $\Gamma \sim 1.01$ in order for the specific internal energy to be positive. 
Once a star is relaxed to the desired density profile, we apply a modification step where we assign the correct values for $P$ and $u$ for a given $\rho(\vec{r})$ and nuclear EoS.
After that, we can dynamically evolve the star. 

\subsection{Spherically symmetric and static neutron star tests}
\label{subsec:tov_tests}
\begin{figure*}
    \centering
    \subfigure[Radial oscillation modes]
    {\includegraphics[width=0.47\textwidth]{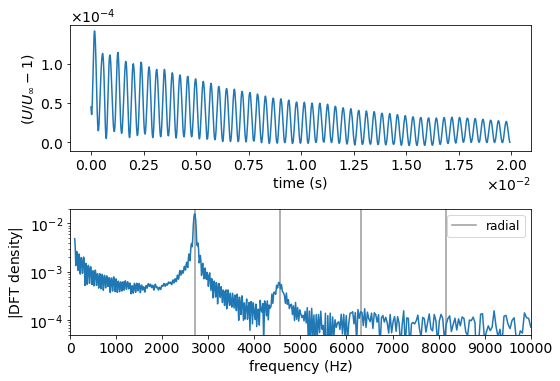}
    \label{fig:osc_radial}}
    \hfil
    \subfigure[Non-radial modes: $l=1, m=0$]
    {\includegraphics[width=0.47\textwidth]{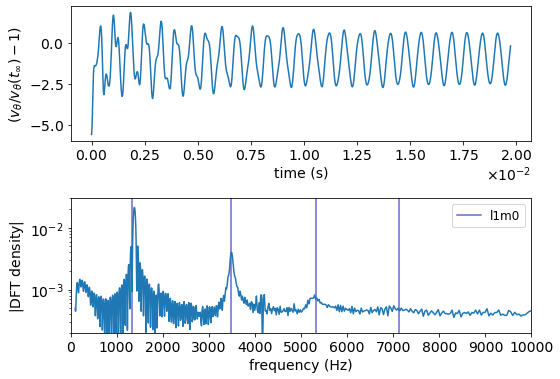}
    \label{fig:osc_l1m0}}
    \subfigure[Non-radial modes: $l=2, m=0$]
    {\includegraphics[width=0.47\textwidth]{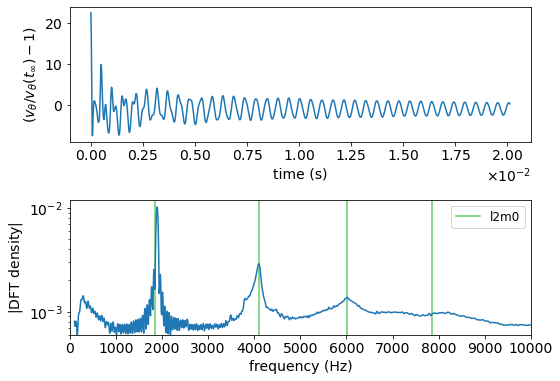}
    \label{fig:osc_l2m0}}
    \hfil
    \subfigure[Non-radial modes: $l=3, m=0$]
    {\includegraphics[width=0.47\textwidth]{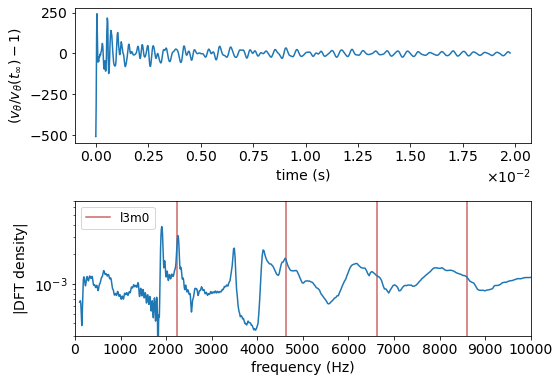}
    \label{fig:osc_l3m0}}
    \caption{The four subplots demonstrate the oscillations of the radial and $l=1,2,3$ modes of a TOV neutron star. For each subplot, the top panel shows the oscillation of internal energy for radial modes and $v_\theta$ for non-radial modes. The bottom panel is the frequency spectrum in comparison to the modes by Font et al. with a 1D grid-based code~\cite{font00}.}
        \label{fig:osc_tov}
\end{figure*}

To verify the implementation of relativistic SPH for neutron stars, we apply velocity perturbations to a star (initially in equilibrium) and extract the frequency modes of the resulting oscillations. 
The frequencies are compared to previous work which used a 1D grid-based code~\cite{font00, font01}.

We use a polytropic EoS with $K=100$ \edit{}{in geometric units ($1.46\times10^5$ in cgs units)} and $\Gamma=2$ and initialize a star with a central density of $\rho = 0.00128$ in geometric units (or $7.9\times 10^{14} \: \mathrm{g/ {cm}^3}$ for mass unit $M = M_\odot$). 
Solving the TOV equations results in a star with total mass of 1.4 solar masses and radius of $R=9.59$ geometric units (or $14.2 \: \mathrm{km}$). 
We use approximately 15,000 particles.
After relaxing the neutron star in an external potential and evolving it to equilibrium in the fixed GR background, we excite different oscillation modes. 
For radial modes, a small Gaussian perturbation is introduced to the radial particle velocity $v_r$. 
For non-radial modes, the added perturbation is 
\begin{align}
        (\delta v_\theta)_{nlm} &=  A \nabla_\theta\left( \sin\left(\frac{n\pi}{R_0} r\right)Y_l^m(\theta,\phi) \right) \nonumber\\
        &= A \sin\left(\frac{n\pi}{R_0} r\right)e^{i m \phi} P_l^{(m+1)}\cos{\theta}\,,
\end{align}
where $R_0$ is the star's radius, $Y_l^m$ are the spherical harmonics, and $P_l^m$ are the associated Legendre polynomials.
To extract radial oscillation modes, we analyze the internal energy since it shares the same oscillation patterns as $v_r$.
For non-radial modes, we use $v_\theta$.
For the latter, we \edit{use}{take a slab of} particles at ${r=(0.5\pm0.04) R_0}$ \edit{with a bandwidth of $0.04 R_0$}{} and $\theta = 2\pi/3 \pm 0.08\pi$\edit{ with a bandwidth of $0.08 \pi$}{}. 
The velocity of the selected particles is then described by a linear fit in the region in $r$ and in $\theta$.

The results are shown in Fig.~\ref{fig:osc_tov}. 
Overall, we find that we can reproduce the expected two to three lowest frequencies for radial and non-radial oscillation modes very well~\cite{font00,font01}. 
Higher frequencies might only be accessible with larger particle numbers. 
For the non-radial perturbation with $l=3, m=0$, we find that a mix of oscillation modes with $l=3$, $2$, and $1$ is excited. 
We suspect that this might be either due to limited resolution or the small initial overlap with the $l=1,2$ modes that become noticeable as the $l=3$ mode dampens out faster with viscosity.
However, the overall agreement with previous work is encouraging so that we proceed to simulate stars in fixed GR with a solid crust. 

\section{Neutron stars with a solid crust in SPH}
When fully exploring the deformation and potential shattering of the neutron star crust, we require the information on its shear modulus $\mu$, yield stress $Y_0$, and breaking strain.
These quantities can in principle be determined from microscopic calculations using the Wigner-Seitz approximation or e.g. Molecular Dynamics simulations of neutron star crust material~\cite{Horowitz2009,Caplan2018}. 
In the current work, we focus on elastic deformations and, as a consequence, only require information on $\mu$.
The shear modulus of the neutron star crust is given by Coulomb interactions and depends on density and composition.
\begin{figure}
    \centering
    {\includegraphics[width=0.45\textwidth]{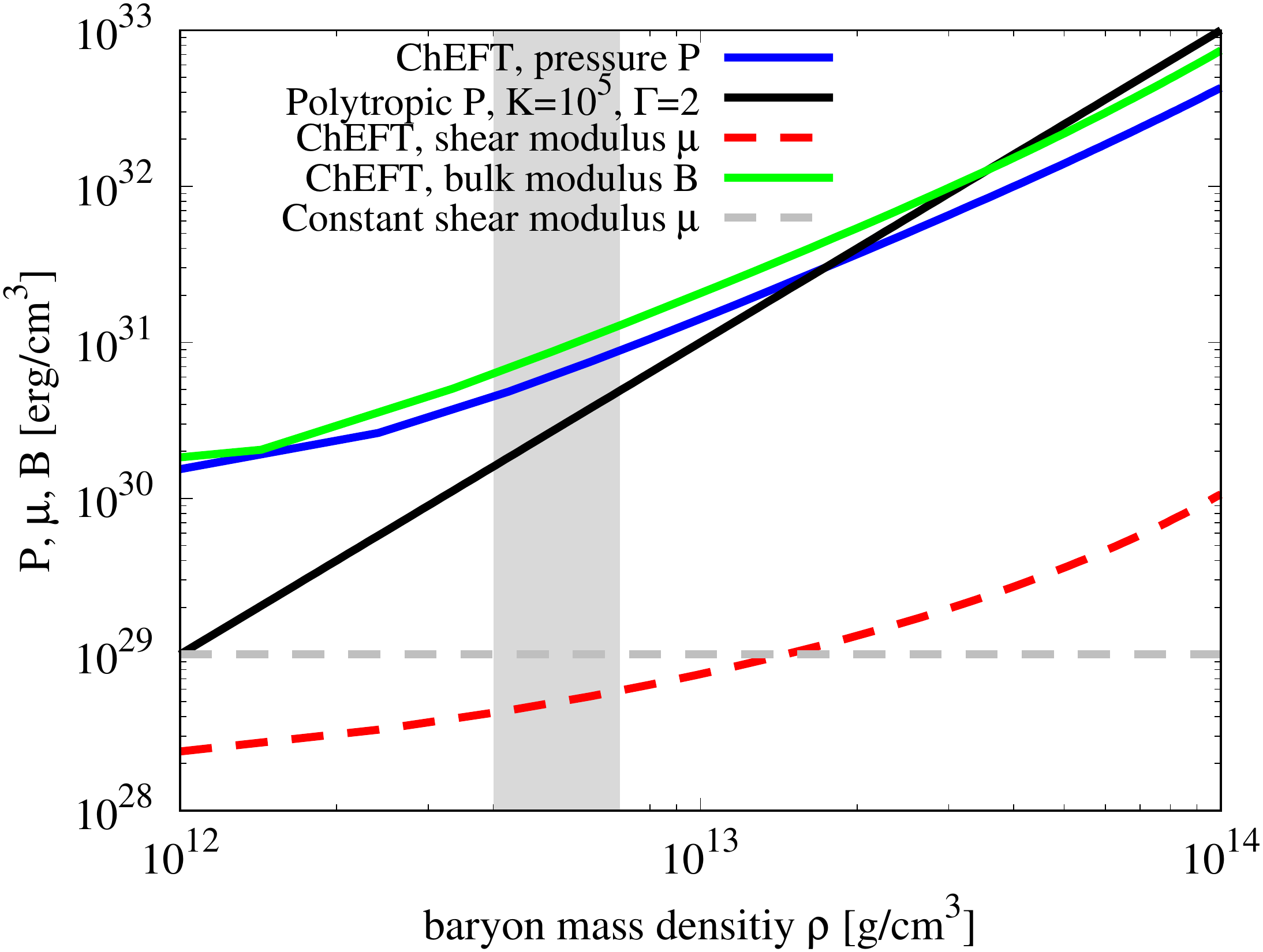}
    \caption{Pressure $P$ in the neutron star crust as a function of baryon mass density $\rho$ using a polytropic EoS that is used throughout the paper together with a constant shear modulus $\mu = 10^{29} \: \mathrm{erg/cm^3}$. For comparison, we also plot $P$, $\mu$, and the bulk modulus $B$ calculated using nuclear interactions from Chiral Effective Field Theory in Ref.~\cite{Tews:2018kmu}. The grey band shows densities $4\times 10^{12} \: \mathrm{g/cm^3} \leq \rho \leq 7 \times 10^{12} \: \mathrm{g/cm^3}$ corresponding to the crust setup in section~\ref{sec:solid_crust_osci}.}
    \label{fig:cheft_all}}
\end{figure}
In the Wigner-Seitz approximation the periodic structure of the lattice is described by using a unit cell which contains one lattice point. 
To describe the inner crust, neutron matter is uniformly distributed within the spherical cell with radius $R_W$ and volume $V_W$.
A spherical nucleus with radius $R_0$, volume $V_0$, and proton number $Z$ resides in the center of the cell. 
For such a configuration, the shear modulus can be calculated as 
\begin{align}
    \mu = 0.1194 \: \left( 1 - 0.010 \: Z^{2/3} \right) \: n_i \left(Ze\right)^2 a^{-1}\,,
\end{align}
where $n_i = 1/V_W$ and $a = R_W$~\cite{tews2017}.

Figure~\ref{fig:cheft_all} shows $P$ as a function of baryon mass density $\rho$ as well as $\mu$ and the bulk modulus $B$. The latter is given by the nuclear EoS and can be determined as a function of baryon number density $n_b$ via:
\begin{align}
    B = - V \frac{\partial P}{\partial V} = n_b \frac{\partial P}{\partial n_b} .
    \label{bulk_modulus}
\end{align}
The pressure, shear modulus and bulk modulus are calculated using the Wigner-Seitz approximation~\cite{tews2017} using the chiral effective field theory calculations of ~\cite{Tews:2018kmu}.
The shown range in $\rho$ corresponds to densities in the inner crust with typical values for $\mu$ in the range of $\left(10^{28} - 10^{30}\right) \: \mathrm{erg/cm^3}$. 
With that, the shear modulus in the neutron star crust is much smaller than the bulk modulus, leading to toroidal oscillations of the solid crust to have lower frequencies than fundamental fluid modes. 
At the same time, the smaller values of $\mu$ in comparison to $B$ might lead to challenges when modeling purely toroidal oscillations of the crust since numerical effects of particle bulk motion might arise.  
For comparison, Fig.~\ref{fig:cheft_all} also shows \edit{a frequently used polytropic EoS}{the polytropic EoS as well as a constant $\mu = 10^{29} \: \mathrm{erg/cm^3}$ that are both applied in this paper.} 
Following Eqs.~(\ref{poly_eos}) and (\ref{bulk_modulus}), the bulk modulus for a polytropic EoS is simply given by $B = \Gamma P$.

\edit{When setting up neutron stars in flecsph, we}{In the code, we use different particle types to} distinguish between fluid core and solid crust particles.  
\begin{figure}
    \centering
    {\includegraphics[width=0.37\textwidth]{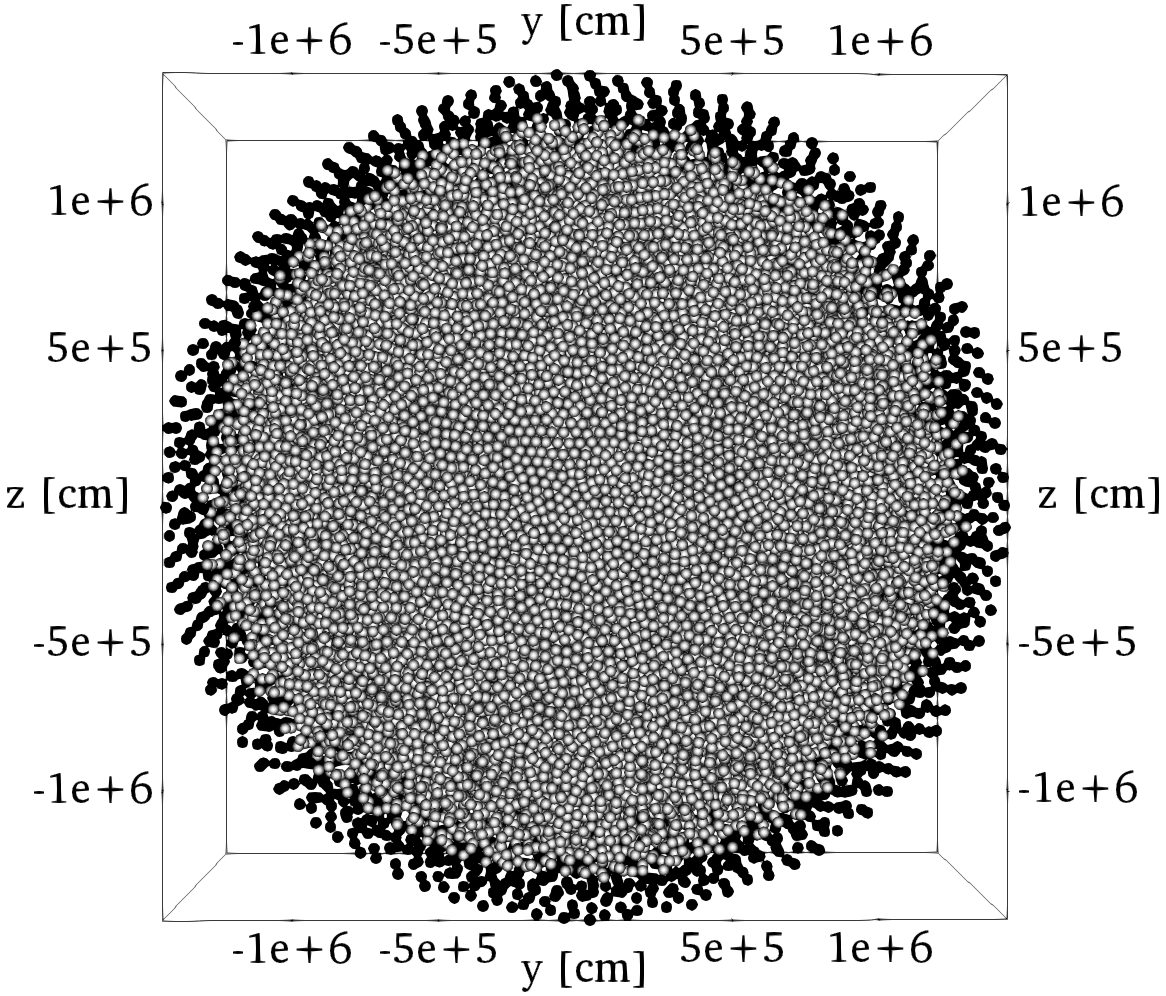}
    \caption{Cut through a 1.4 solar mass neutron star with a solid crust (black points) and fluid core (gray points) set up with a polytropic equation of state.}
    \label{fig:star_crust}}
\end{figure}
The \edit{first}{former} are evolved according to the hydrodynamic conservation equations with fixed GR metric while for the \edit{crust particles}{latter} we add the solid components from Eqs.~(\ref{solid1})~-~(\ref{solid4}).
We want to point out that in this work, we combine Newtonian elastodynamics terms with relativistic hydrodynamics in fixed GR background. 
This simple approach has to be improved in the future. 
As a next step, we therefore plan to apply the Lie derivative instead of the Lagrangian time derivative and take into account
the time lapse and the coordinate shift. 
This is equivalent to using Newtonian elastodynamics in locally flat space (justifiable by the relatively calm spacetime curvature around a neutron star) and transforming the resulting accelerations to the global coordinate frame.
A fully consistent approach with curvature effects entering at the level of computing the deviatoric stress tensor is planned as the final step in our work. 

Modeling the dynamics of the solid neutron star crust in SPH comes with several challenges. 
In general, the stability of a neutron star is the result of the gravitational (compression) and nuclear (repulsion) forces being in balance such that the dynamics of the crust and core are tightly coupled for e.g. radial oscillations.
The situation is different for shear deformations. 
Neutron star cores are composed of superfluid nuclear matter and therefore do not sustain shear forces.
The crystal material in the crust, on the other hand, provides restoring forces to shear deformations. 
With that, when studying e.g. toroidal motion in neutron stars and neglecting effects of the magnetic field, the dynamics of the crust and the core should be decoupled. 
For this, advanced models for the numerical viscosity are required. 
In addition, free-slip boundary condition at the bottom of the crust can filter neighbor interactions including modifications to the stress tensors to remove shear stress across the crust-core interface.
However, the latter might not be easy to identify. 
Although the crust-core transition density is well defined by nuclear physics, for neutron stars modeled with SPH it does not necessarily correspond to a sharp interface. 
This can be seen in Fig.~\ref{fig:star_crust} where we plot a neutron star for about $10^5$ crust (black points) and core (gray points) particles. 
Moreover, \edit{for example for}{as an example, for} tidally deformed stars the interface is not a static spherical boundary and has to be determined numerically at each time step. 

Finally, the finite resolution in the crust can also be a difficulty. 
\edit{From at about $10^{14} \: \mathrm{g/cm^3}$ the crust extends}{The crust extends from the density of about $10^{14} \: \mathrm{g/cm^3}$}
down to the surface of the star. 
While the radial extent is about 1-2km (ca. 10\% of the neutron star radius), the covered density range spans about 14 orders of magnitude~\cite{Chamel2008}. 
It is needless to say that using equal-mass SPH particles it is impossible to resolve the entire crust. 
The question is whether nuclear matter at lowest densities is crucial for the dynamics of the solid crust.  
As shown in Fig.~\ref{fig:cheft_all}, the crustal shear modulus increases with density. 
It is largest in the inner crust, \edit{before it falls off}{right before it vanishes} at the crust-core interface. 
We can therefore expect that the inner crust provides the majority of the restoring force for shear and tangential deformations. 
Furthermore, the i-mode is peaked at the base of the crust~\cite{Neill2021}. 
During oscillations, the core is ellipsoidally deformed, while the crust shears away from the bulge. 
This nature of motion, as bulk motion of the core vs. bulk motion of the crust, most likely does not rely on resolving the low density crust.

\edit{Nevertheless, we are currently pursuing two methods}{Two approaches can be taken} to address the crust-core decoupling and the finite density resolution in the crust.  
The first approach relies on replacing the neutron star core by an external potential $\phi_\mathrm{eff}$ and potential walls. 
\begin{figure}
    \centering
    {\includegraphics[width=0.3\textwidth]{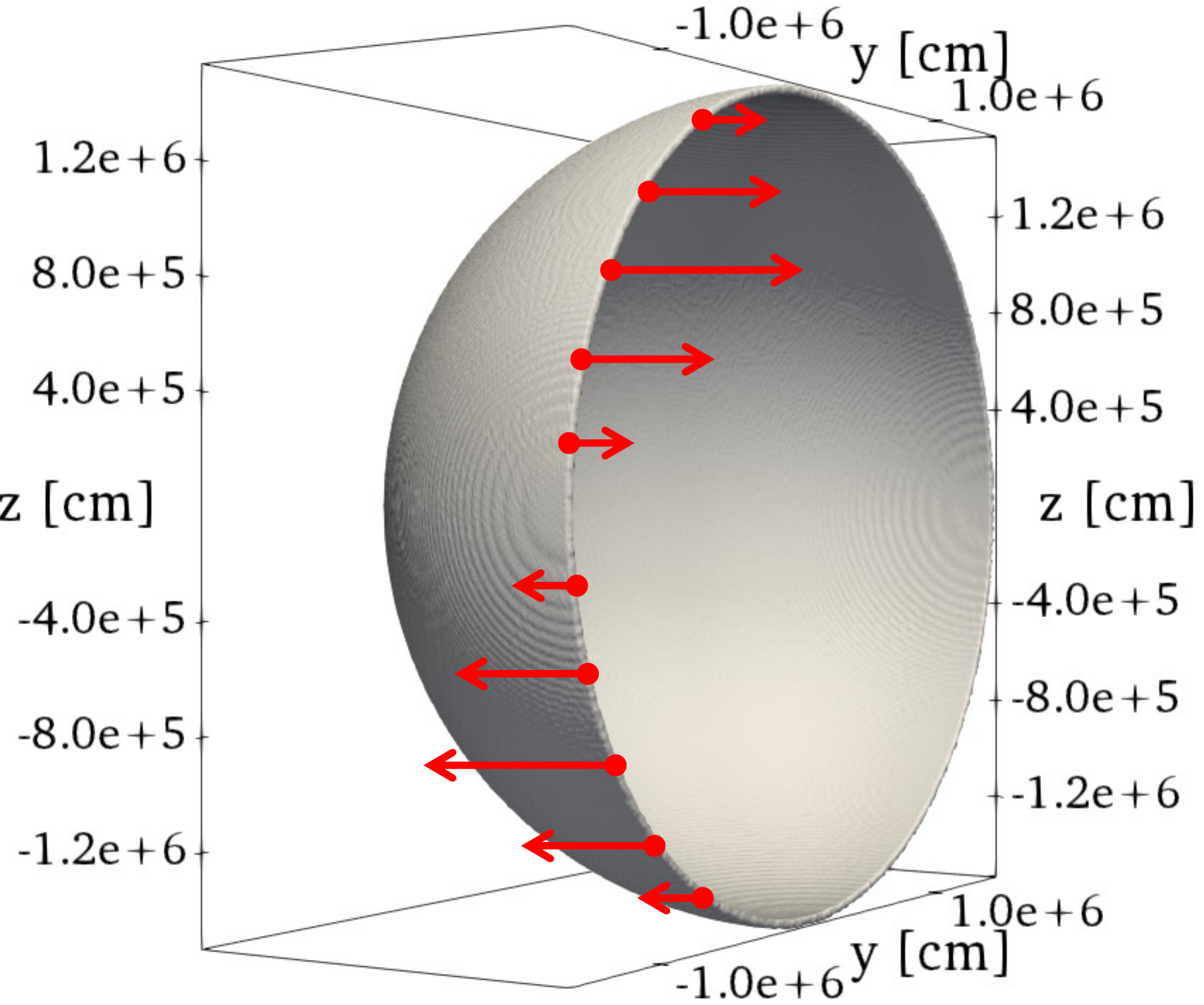}
    \caption{A shell of neutron star crust with densities $4\times 10^{12} \: \mathrm{g/cm^3} \leq \rho \leq 7 \times 10^{12} \: \mathrm{g/cm^3}$ using external potentials to hold the crust particles in place. Arrows show the imposed toroidal velocity in the crust.}
    \label{fig:star_shell}}
\end{figure}
The idea is to use the effective potential to ensure that the particle distribution follows a given density profile in the crust, while potential walls hold the crust in place. 
By selecting the radial location of the walls, we can model different low-density regions of the crust which might otherwise not be accessible when modeling the entire star. 
\edit{At present, this method is only available for simple setups, e.g. isolated and spherically symmetric neutron stars.
However, an extension to e.g. tidally deformed stars is possible. }{This approach enables the study of the crust behavior in isolation, and admits not only the simplest setups, e.g. isolated and spherically symmetric configurations, but also tidally-deformed and time-varying configurations.}
Figure~\ref{fig:star_shell} shows an example of such a crust simulation. 
Here, one million SPH particles have been used to simulate the crust for densities $4 \times 10^{12} \: \mathrm{g/cm^3} \leq \rho \leq 7 \times 10^{12} \: \mathrm{g/cm^3}$. 
\begin{figure}
    \centering
    {\includegraphics[width=0.48\textwidth]{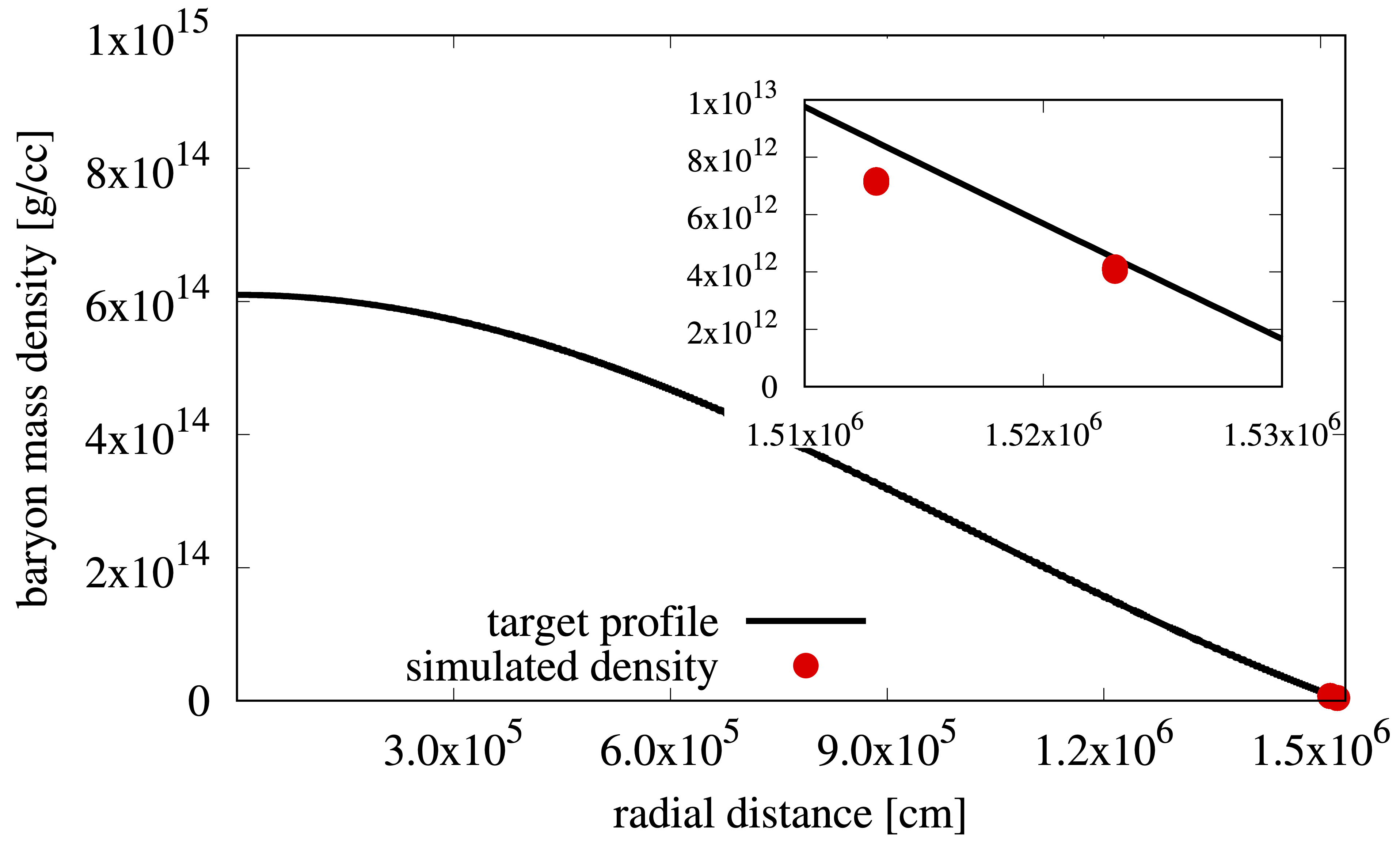}
    \caption{Density of a 1.4 solar mass neutron star from the center to its surface at radial distance $R \sim 1.54\times 10^6 \mathrm{cm}$. The simulated crust in Fig.~\ref{fig:star_shell} with corresponding densities is marked by red points which show the position of all one million SPH particles.} 
    \label{fig:star_rho_shell}}
\end{figure}
Despite the large particle number, the crust is represented by only two layers\edit{ which location is}{, as} shown in Fig.~\ref{fig:star_rho_shell}. 
Due to the small width of this shell, the density determination is off by about (10-20)\%. 
An SPH approach that treats the shell like a 2D object, especially for the density determination, might improve the numerical representation.  
However, this result is still very encouraging considering that $\rho \sim 10^{12} \: \mathrm{g/cm^3}$ would probably not be accessible when modeling the entire star and that we are resolving a radial width of $< 1\%$ of the total radius.  
We used this setup for a toroidal oscillation study of the neutron star crust which will be discussed in the next section. 
The effective potential approach has been used for neutron stars without a fixed GR background and \edit{, as a next step, will be ported}{can also be extended} to model relativistic setups. 
Another approach that is currently under development models the entire star but represents the crust by a \edit{one-layer outer}{single-layer particle} shell on the neutron star surface. 
Neighbor interactions between crust and core particles are filtered and modified to eliminate shear stresses.
The crust is described as a 2D object with radially integrated crust properties so that the radial density structure does not have to be resolved.  

\section{Neutron Star Oscillations with a solid crust}
\label{sec:solid_crust_osci}
A simple test of the radial oscillation mode similar to the one performed in Sec.~\ref{subsec:tov_tests} can be demonstrated with a neutron star with a solid crust. 
As expected, we found the results with solid and fluid crusts to be identical. 
Radial stellar compression and decompression should not be affected by shear forces and even if numerical artefacts result in shear forces for individual particles, the effects should be dominated by the pressure due to the previously discussed difference in $\mu$ and $B$.

Using the toroidal oscillation setup in the crust layer as given in Fig.~\ref{fig:star_shell}, we perform dynamical simulations and compare the results to a crust which is assumed to be fluid. 
We use a polytropic EoS as given in Fig.~\ref{fig:cheft_all} and a constant shear modulus of $\mu = 10^{29} \mathrm{erg/cc}$.
\begin{figure}
    \centering
    {\includegraphics[width=0.48\textwidth]{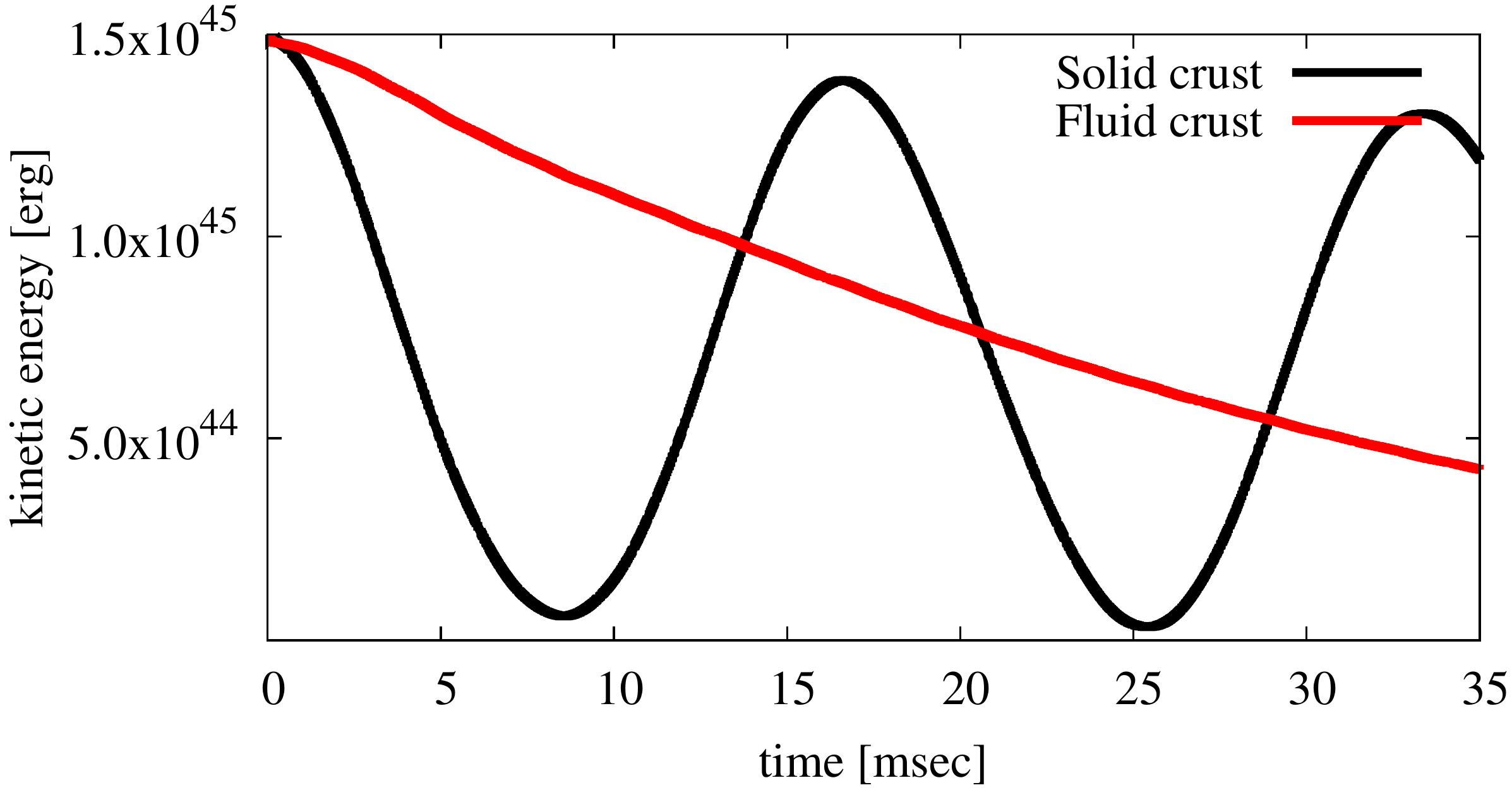}
    \caption{Kinetic energy of the toroidally oscillating crust in Fig.~\ref{fig:star_shell} with a solid and fluid treatment.} 
    \label{fig:osci_crust}}
\end{figure}
The kinetic energy as a function of time is shown in Fig.~\ref{fig:osci_crust}. 
We can see an oscillation pattern in the solid crust while for the fluid crust the initially setup toroidal velocity decays with time due to numerical viscosity. 
The resulting oscillation frequency of velocity for the solid crust is about $f_\mathrm{cal} = 30\: \mathrm{Hz}$.
When corrected for gravitational redshift, this would lead to an observable frequency of $f_\mathrm{obs} = f_\mathrm{cal} \sqrt{1 - (r_S/R)} \sim 25.18 \: \mathrm{Hz}$ with the Schwarzschild radius $r_S = 2GM/c^2$ where $G$ is the gravitational constant~\cite{tews2017}. 
As expected, this is much smaller than the fundamental fluid modes discussed in section~\ref{subsec:tov_tests}. 
The value is comparable to frequencies that are extracted from quasiperiodic oscillations of neutron stars after giant X-ray flares (ca. 18~Hz) and are associated with the fundamental toroidal mode of the neutron star crust~\cite{Strohmayer:2005ks, Watts:2005ue}. 
Given the simple approach this result is very encouraging. 
As the next step, this study should be repeated with more realistic EoS and shear modulus in a fixed GR background. 


\section{Conclusion}
We present first results on our ongoing work to simulate the dynamics of the solid neutron star crust in the general relativistic (GR) regime using a fixed background metric. 
Our current work allows us to generate TOV neutron stars and study radial and non-radial oscillations modes.
The resulting oscillation frequencies show very good agreement with published works. 
The simulation of the solid neutron star crust comes with several challenges, for example, its large density range that spans fourteen orders of magnitude and the requirement to decouple shear interactions between the crust and the core. 
Using external potentials, we can simulate desired density ranges in the crust without the need to include the core. 
Preliminary studies of toroidal crust oscillations show good agreement with published work but need to be improved in the future including more accurate models for the nuclear equation of state and shear modulus. 
For radial oscillation modes of the entire star in a fixed GR background metric, we do not find changes when including the solid crust, which agrees with expectations. 
\edit{For future follow-ups, the implementation of spinning neutron stars and fully-dynamical general relativity background will be needed to more accurately simulate binary neutron stars. 
Furthermore, the elimination of the tangential forces passing through the crust-core interface and scaling up the simulation to larger particle number to better resolve the crust will be useful in fully capturing the crust dynamics during the neutron star merger.}{Finally, realistic simulations of binary neutron star mergers will require the usage of a dynamical general relativistic metric. This is left for future work. }

\section*{Acknowledgments}
\edit{}{We would like to thank Stephan Rosswog for helpful comments.}
The presented research was supported by the Advanced Simulation and Computing program and the Laboratory Directed Research and Development program of Los Alamos National Laboratory (LANL) under project numbers 20190021DR and 20200145ER. 
This research also resources provided by the LANL Institutional Computing Program, which is operated by Triad National Security, LLC, for the U.S. Department of Energy National Nuclear Security Administration under Contract No. 89233218CNA000001. This work is authorized for unlimited release under LA-UR-21-24191.

\bibliographystyle{IEEEtran.bst}
\bibliography{IEEEabrv,refs}

\end{document}